\documentclass{aa}
\usepackage{epsfig}
\begin{document}
   \title{Orbital and spin variability of the Intermediate Polar BG CMi}
   \author{Y.G. Kim\inst{1,}\inst{2} \and I.L. Andronov\inst{3,}\inst{4,}\inst{1}  \and S.S. Park\inst{1} \and Y.-B. Jeon\inst{5}
         }
   \offprints{I. L. Andronov, {\em e-mail: il-a@mail.ru}}
   \institute{
University Observatory, Chungbuk National University, 361-763,
Cheongju, Korea
         \and
Institute for Basic Science Research, Chungbuk National
University, 361­763, Korea
         \and
Department of Astronomy, Odessa National University,
             T.G.Shevchenko Park, 65014, Odessa, Ukraine,
         \and
         Crimean Astrophysical Observatory,
         98409 Nauchny, Ukraine,
         \and
Korea Astronomy Observatory and Space Science Institute, Daejeon 305-348, Korea
             }
\date{Received March 11, 2005 / Accepted June 14, 2005}
\abstract{Results of a CCD study of the variability of the
cataclysmic variable BG CMi obtained at the Korean 1.8m telescope in
2002-2005 are presented. The "multi-comparison star" method had
been applied for better accuracy estimates. The linear ephemeris based on 19 mean maxima for 2002--2005  is HJD $2453105.31448(6)+0.01057257716(198)(E-764707).$ The period differs from that predicted by the quadratic ephemeris by Pych et al. (1996) leading to a possible cycle miscount. The statistically optimal ephemeris is a fourth-order polynomial, as a quadratic or even a cubic ephemeris leads to unaceptably large residuals: Min.HJD=$
2445020.28095(28)+0.0105729609(57)E
-1.58(32)\cdot10^{-13}E^2-5.81(64)\cdot10^{-19}E^3+4.92(41)\cdot10^{-25}E^4.$
Thus the rate of the spin-up of the white dwarf is decreasing. An alternative explanation is that the spin-up has been stopped during recent years.
The deviations between the amplutudes of the spin variability in V and R, as well as between phases are not statistically significant. However, the orbital light curves exhibit distinct difference; the corresponding color index shows a nearly sinusoidal shape with a maximum at orbital phase $\sim0.2.$ The variations of the amplitude of spin waves shows a short maximum at the phase of the orbital dip. The corrected
ephemeris for orbital minima is
Min.HJD=$2448368.7225(12)+0\fd13474841(6)\cdot(E-24849)$
 with a narrow dip occuring
0.07P later. The rate of the spin period variation seems to be
changed, justifying the necessity of regular observations of intermediate polars.
 }
\authorrunning{Kim, Andronov, Park, Jeon}
\titlerunning{Orbital and Spin Variability of BG CMi}
\maketitle


\section{Introduction}
The object 3A 0729 + 103 has been identified as an intermediate
polar-type cataclysmic variable by McHardy et al. (1984). They
found night-to-night variations of the mean brightness with an
amplitude up to 0\fm22 in their sample. The orbital period was
found to be 3$^{\rm h}$14\fm1 and maximal amplitude up to 0\fm15.
However, the most striking feature is the 15.2 min periodic
variations of variable shape, which may be interpreted as the spin
period of a magnetic white dwarf. In the General Catalogue of Variable Stars, the designation
of the star is BG CMi.

The general model for intermediate polars is a red dwarf filling its
Roche lobe, and a white dwarf, the magnetic field of which is
strong enough to disrupt accretion disk completely or at least in
its internal parts (cf. Patterson 1994, Warner 1995, Hellier
2001). Norton et al. (1992) suggested that the 913 sec period is a
spin-orbital beat period, thus indicating a weak or absent
accretion disk and a fast flipping of the stream from one active
pole to another.

Patterson and Thomas (1993) suggested the double period of 1827
sec instead of the observed 913 sec. Such a model corresponds to
nearly equal accreting columns lying near the equatorial plane.

More recently, de Martino et al. (1995) concluded that this
period is the spin one. From a Doppler tomogram analysis, Hellier
(1999) reported on a ``weak spin wave" with the same period.

The circular polarization has been detected by Penning et al. (1986).
However, no evidence on its modulation with spin period has been
detected, contrary to e.g. V405 Aqr (Shakhovskoy and Kolesnikov 1997),
which could justify the suggested spin period value.

The 15 min variations show a drastic period decrease, as was
originally found by Augusteijn et al. (1991) and Singh et al. (1991),
and studied in more detail by Patterson and Thomas (1993) and Pych
et al. (1996). Garlick et al. (1994) even suggested a cubic term
in the ephemeris.

The study of rotational evolution of the white dwarf needs
continuous monitoring of such objects. In the present paper we
report on further studies of the variability of the star based on the
CCD observations obtained in the V and R filters at the 1.8m
telescope in 2002-2005.

\section{Observations and comparison stars}
The observations have been obtained with a thinned SITe 2k CCD
camera attached to the 1.8m telescope of the Korea Astronomy Observatory and Space Science Institute (Bohyunsan).
The instrumental V and R systems have been used. To
determine instrumental magnitudes of stars, the IRAF/DAOPHOT package
(Massey \& Davis 1992) has been used.

Altogether, 1\,236 V and 170 R observations have been obtained during 9 nights from December
26, 2002 to February 3, 2005 (JD 2452635-3405) with an
integration time 15$^{\rm s}$ in 2002 and 30$^{\rm s}$ in 2004 and 2005.

The journal of observations is presented in Table 1. In an
addition to the mean magnitudes $\langle m\rangle$ in the
instrumental system m (see description below), we list the r.m.s.
deviations $\sigma_m$ from the mean as an estimate of the
characteristic total amplitude consisting of the variability at
a few time scales and of the observational noise.

For further analysis of the individual runs, we mark them by the
filter letter and five last digits of the integer part of the
Helicentric Julian Date (HJD) of the beginning of the run.


\subsection{Using few comparison stars}

\begin{figure}
\label{f1}
\psfig{file=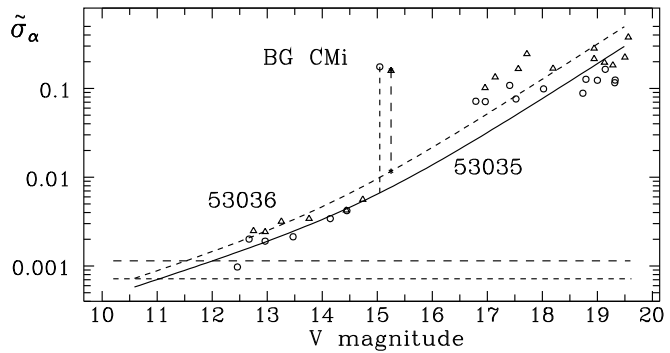}
\caption{The dependence on brightness of the r.m.s. deviation of stellar magnitudes of the variable and the comparison stars on the corresponding sample mean values. Open circles - for the night 53035 with relatively good weather, open triangles - for the night 53026 with highly variable atmospheric transparency. The curved lines correspond to the best fit assuming the Poisson noise for the counts of the star and background. Horizontal lines correspond to the estimated accuracy of the ``artificial" mean weighted comparison star. The bottom curve in these two pairs correspond to better atmospheric conditions and thus smaller error estimates.
Please note that the brightness is calibrated using results from the night 53035, thus the points for the night 53036 with worse weather are shifted towards larger apparent magnitude and larger error estimates. Vertical lines project mean magnitude of BG CMi onto corresponding "accuracy-brightness" dependence.}
\end{figure}

\begin{table}
\label{t1}
\caption{Journal of observations: begin and end of the run, the number
of observations $n,$ the mean values $\langle m\rangle$ and r.m.s. deviations of the mean $\sigma_m$ and the filter.}
\begin{tabular}{lcrrcc}
\hline
$t_{start},$ 24.....&$t_{end},$ 24.....&\hfill$n$\hfill&$\hfill\langle m\rangle$\hfill&$\sigma_m$&Filter\\
\hline
52635.2721& 52635.3388&~~95& 15.052& 0.147&V\\
52638.2627& 52638.3379&~177& 14.941& 0.137&V\\
53035.0390& 53035.1655&~217& 15.042& 0.176&V\\
53036.1058& 53036.2677&~278& 14.953& 0.163&V\\
53052.9374& 53053.1193&~299& 15.023& 0.171&V\\
53383.0829& 53383.1145&~~18& 15.046& 0.147&V\\
53383.0838& 53383.1153&~~17& ~2.800& 0.149&R\\
53384.1352& 53384.1648&~~14& 14.914& 0.173&V\\
53384.1365& 53384.1657&~~14& ~2.678& 0.171&R\\
53385.9945& 53386.1484&~~66& 15.049& 0.193&V\\
53385.9955& 53386.1494&~~69& ~2.811& 0.184&R\\
53404.9714& 53405.1264&~~72& 15.077& 0.206&V\\
53404.9726& 53405.1250&~~70& ~2.848& 0.197&R\\
52635.2721& 53405.1264&1236& 15.005& 0.173&all V\\
53383.0838& 53405.1250&~170& ~2.814& 0.189&all R\\
\hline
\end{tabular}
\end{table}

\begin{table*}
\label{t2}
\begin{center}
\caption{Characteristics of the multi-sinusoidal fit of individual
runs of observations of BG CMi. The flux is expressed in units of
$10^{-15}$ erg s$^{-1}$cm$^{-2}$\AA$^{-1}.$ The calibration in V
has been made using the magnitude of the comparison star. For R,
the standard magnitude is unknown and has been arbitrarily set to
zero.This makes values of intensity to be apparently large. They
may be calibrated after further determination of the R magnitude
of the comparison star. For bad phase distribution caused by an
incomplete coverage of the period in the first two runs, the
values are highly biased and should not be taken for further
analysis. Such biased or statistically insignificant values are
marked by a semi-column ":".}
\begin{tabular}{cccccccc}
\hline
Date&$I_0$&$R_1$&$R_2$&$R_3$&$R_4$&$R_5$&$R_6$\\
\hline
V 52635&-759: $\pm$ 129 &1381: $\pm$ 2318&1018: $\pm$ 1670& 599: $\pm$  945& 270: $\pm$  400&  85: $\pm$  115&  14: $\pm$  17 \\
V 52638& 46:  $\pm$49   &77:   $\pm$89   & 60:  $\pm$67   & 39:  $\pm$41   & 20:  $\pm$20   &7.17: $\pm$6.67 &1.39: $\pm$1.29 \\
V 53035&3.527 $\pm$0.042&0.294 $\pm$0.055&0.337 $\pm$0.054&0.273 $\pm$0.055&0.071:$\pm$0.059&0.050:$\pm$0.059&0.092:$\pm$0.056\\
V 53036&3.796 $\pm$0.035&0.362 $\pm$0.050&0.343 $\pm$0.049&0.167 $\pm$0.050&0.040:$\pm$0.049&0.075:$\pm$0.049&0.062:$\pm$0.049\\
V 53052&3.578 $\pm$0.035&0.442 $\pm$0.051&0.179 $\pm$0.050&0.048:$\pm$0.051&0.068:$\pm$0.049&0.054:$\pm$0.047&0.097:$\pm$0.048\\
V 53385&3.467 $\pm$0.100&0.257:$\pm$0.158&0.296:$\pm$0.158&0.296:$\pm$0.153&0.131:$\pm$0.152&0.059:$\pm$0.136&0.011:$\pm$0.143\\
R 53385&272816$\pm$6747&31239 $\pm$10432&17926:$\pm$10145&20531:$\pm$9874&4393:$\pm$9968&6942:$\pm$9139&1100:$\pm$9558\\
V 53404&3.494 $\pm$0.086&0.406 $\pm$0.124&0.421 $\pm$0.120&0.072:$\pm$0.121&0.135:$\pm$0.121&0.100:$\pm$0.120&0.075:$\pm$0.120\\
R 53404&272413$\pm$6526&31228 $\pm$9288&32071 $\pm$9066&7256:$\pm$9077&6246:$\pm$9273&5251:$\pm$9189&3509:$\pm$9264\\
\hline
\end{tabular}
\end{center}
\end{table*}

To obtain better accuracy, we have used few comparison stars in the field.
Their BV magnitudes have been estimated by Henden and Honeycutt (1995).

The procedure of the ``artificial" mean weighted star has been
used, which has been described by Andronov and Baklanov (2004) and
applied to the comparison stars in the vicinity of BG CMi by Kim
et al. (2004). The instrumental magnitudes of all comparison stars
have been checked, and the outstanding values have been excluded
from the analysis. From the images with all ``good" measurements
of the comparison stars, the mean magnitude differences have been
computed for all comparison stars, then the magnitudes of all
comparison stars have been determined using the magnitude of the
``main" comparison star ``05" (Henden \& Honeycutt 1995)
V=12\fm457, B-V=0\fm707. This star is also marked as "Comp 1" by Pych et al
(1996). Then the weights have been determined for each comparison
star using iterations as described by Kim et al. (2004). Then, for
each image, the brightness of the comparison star has been
determined as the mean weighted value from estimates based on the
magnitude differences between the variable and all comparison
stars.

In Fig. 1, the "accuracy-magnitude" dependence is shown for two
nights with ``good" and ``worse" atmospheric conditions. The
corresponding error estimates for the ``artificial" comparison
star are 0\fm00072 and 0\fm00114. The error estimates of the
variable star itself for these nights may be estimated from a best
non-linear fit as 0\fm0067 and 0\fm0117, respectively.

The observed characteristics of 17 comparison stars are listed in
Kim et al. (2004), the corresponding finding chart is available at
{http://uavso.pochta.ru/bgcmi\_2.GIF\,}.
The differences between our instrumental magnitudes $V_{in}$ and the standard ones
$V_{HH}$ (Henden and Honeycutt 1995) are well explained by the
color reduction formula
\begin{equation}
V_{in}=V_{HH}-0\fm0025(16)+0.(11)\Delta(V-R)_{in}
\end{equation}
(Kim et al. 2004). However, for our photometry, we have not taken
into account the difference between our instrumental VR systems
and the standard ones, as there are no published estimates of the
R magnitudes of the comparison stars. The amplitudes will be
expressed in the instrumental systems without any conversion.

\begin{figure*}
\psfig{file=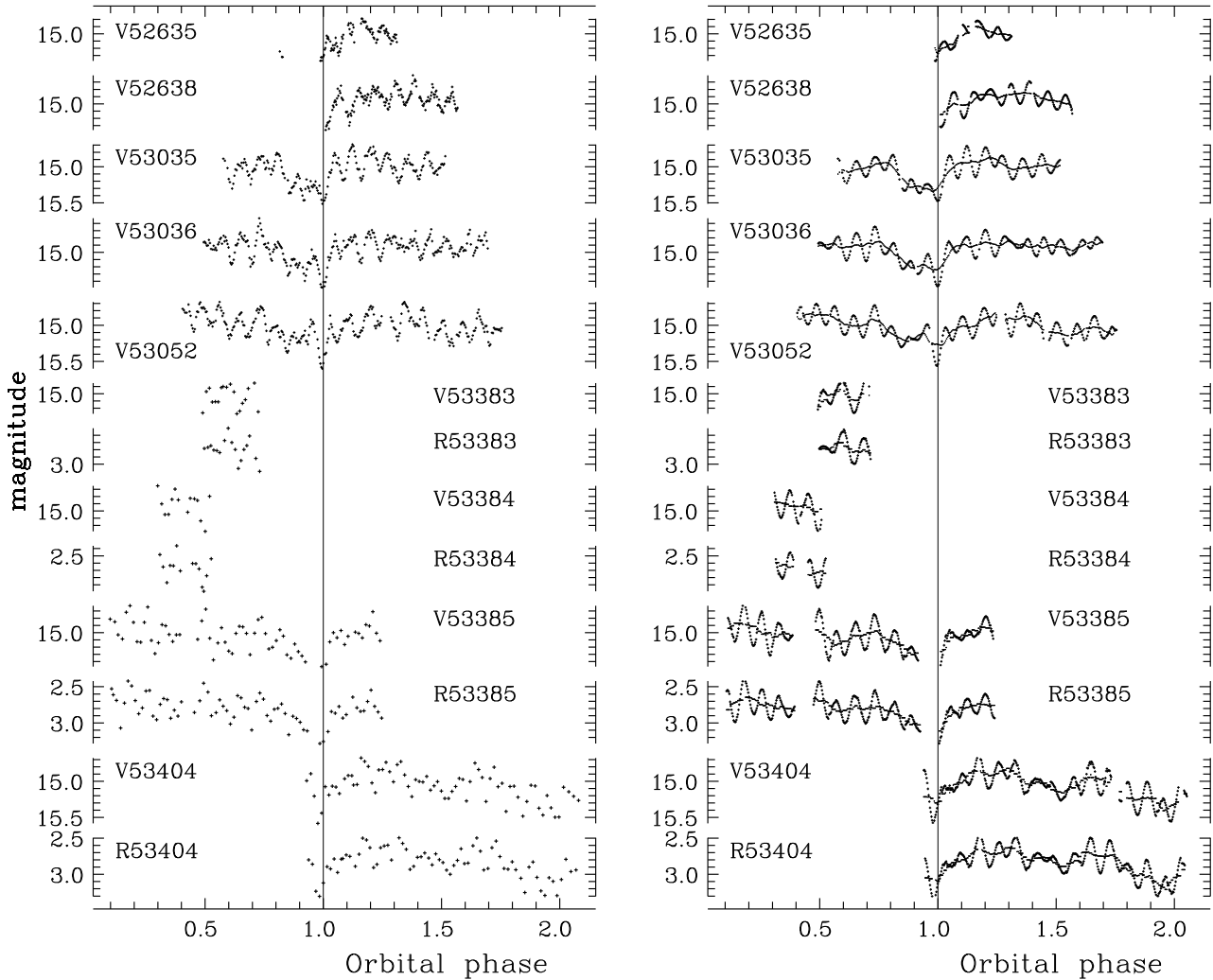}
\label{f2}
\caption{The brightness variations of BG CMi during our observations with orbital phase computed using the ephemeris (4) by Patterson and Thomas (1993).
{\em Left:} original data;
{\em Right:} "running sine" fit $X_{RS}(t,t,\Delta t)$ for times of observations and corresponding local mean $a(t,t,\Delta t).$ In the fits, some points have been skipped because of large error estimates of the smoothing value $(>0\fm10)$ near the gaps of the observations.}
\end{figure*}

The light curves obtained for 9 V runs and 4 R runs are shown in Fig.2.
They exhibit both orbital and spin periodicity, as well as cycle-to-cycle variability, which will be studied below.

\section{Separation of spin and orbital variability}

\subsection{''Running Sine" fit}

The orbital variations show a wave with an asymmetric minimum,
which was interpreted as a grazing eclipse, when the accretion structure is partially eclipsed, but the white dwarf is not (cf. Hellier 1999, 2001). The initial ephemeris
by McHardy et al. (1984) has been revised by Augusteijn et al.
(1991) and Patterson and Thomas (1993). Pych et al. (1993) have
added 5 minima timings and corrected this revised ephemeris.
Patterson and Thomas (1993) distingush the orbital dips, which are
close in phase to the X-ray dips, and the orbital minima
approximately corresponding to the mean between the descending and
ascending branches of the eclipse at it's middle parts. The dip
occurs $\sim0.12P$ after the ``mid-eclipse", as one may see in
their Fig. 6. To study the shape of the orbital curve, they
subtracted the best fitting sinusoid with the spin period for each night.

To take into account the variability of the shape of the 913-sec signal, a local sinusoidal fit
\begin{equation}
X_{RS}(t,t_0,\Delta t)=a-r\cdot\cos(\omega (t-T_0)-2\pi\varphi_0),
\end{equation}
was applied. Here the data at time interval $((t_0-\Delta t),(t_0+\Delta t))$ are taken into account for least squares computation of the coefficients. The parameters $a$ (local mean value), $r$ (semi-amplitude) and $\varphi_0$ (phase of maximum in units of the short period) are dependent both on the mid-time $t_0$ of local sub-interval and the filter half-width $\Delta t.$ Here $\omega=2\pi/P,$ $P$ is period, and $T_0$ is the initial epoch. To follow cycle-to cycle changes, we have adopted the value $\Delta t=0.5P.$

General expressions for the statistical properties of running approximations for arbitrary basic and weight functions have been presented by Andronov (1997), with a comparative study of ``running parabola", ``running sine" and wavelet fits for various weight functions (Andronov 1999). This method had been applied for symbiotic variables UV Aur (Chinarova 1998) and V1329 Cyg (Chochol et al. 1999), where shorter variations with variable shape and nearly constant period are superimposed on (generally aperiodic) long-term trends.

The most recent ephemeris
\begin{equation}
HJD_{Max}=2445020.2800+0.010572992\cdot E
\end{equation}
(Pych et al. 1996) has been used for reference. The long-term variability of the period and thus the initial epoch discussed in next section is negligible for such a {\sl local} fit, the fit obtained for all interval of observations. Thus this initial value of the period may be used for local fits.

As an approximation to the longer-term orbital light curve, we have used local values of $a(t_0,\Delta t)$ at times $t_0$ coinciding with times of real observations. The corresponding values are plotted {\em versus} the orbital phase computed using the ephemeris
\begin{equation}
HJD_{Min}=2445020.384+0.1347486\cdot E
\end{equation}
(Patterson and Thomas 1993). The r.m.s value of accuracy estimate of $a$ is $\sigma_a=0\fm0147$, so total amplitude of variations reaches 22$\sigma_a.$

The corresponding fits for slow variability $a(t,t,\Delta t)$ and
fast+slow variability $X_{RS}(t,t,\Delta t)$ computed for times of
observations, where these fits are available, are shown in the
right part of Fig.2.  One may note a significant shift of the
mid-eclipse as compared to the ephemeris (4).

From our observations, we have determined 3 moments of minima
using the ``running sine" approximation,
which are listed in Table 4 (runs V53035-V53053).
Unfortunately, other runs do not cover eclipses completely,
thus this method was not usable. For those runs (52669-52673, 53386, 53404), we have applied
another method based on multi-sinusoidal fits, which will be discussed
below.

\subsection{Multi-sinusoidal fits}


\begin{figure}
\psfig{file=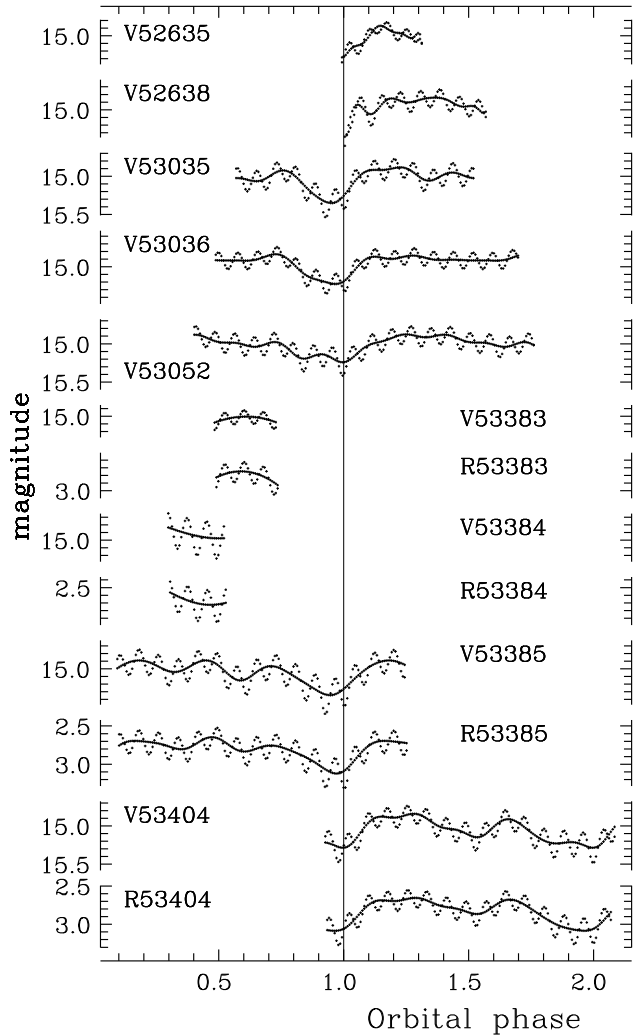}
\label{f3}
\caption{The multi-sinusoidal fit for
individual runs of observations of BG CMi. The orbital light curve
is shown as an abridged sum $(s=6),$ whereas the "orbital+spin"
variations are fitted by a complete sum ($s=7).$ The fits have
been computed for fluxes, following de Martino et al. (1995), and
converted to stellar magnitudes for comparison with the original
data and the "running sine" fits.
For R, the magnitude is expressed as the ``Var-Comp" difference.}
\end{figure}

\begin{figure}
\psfig{file=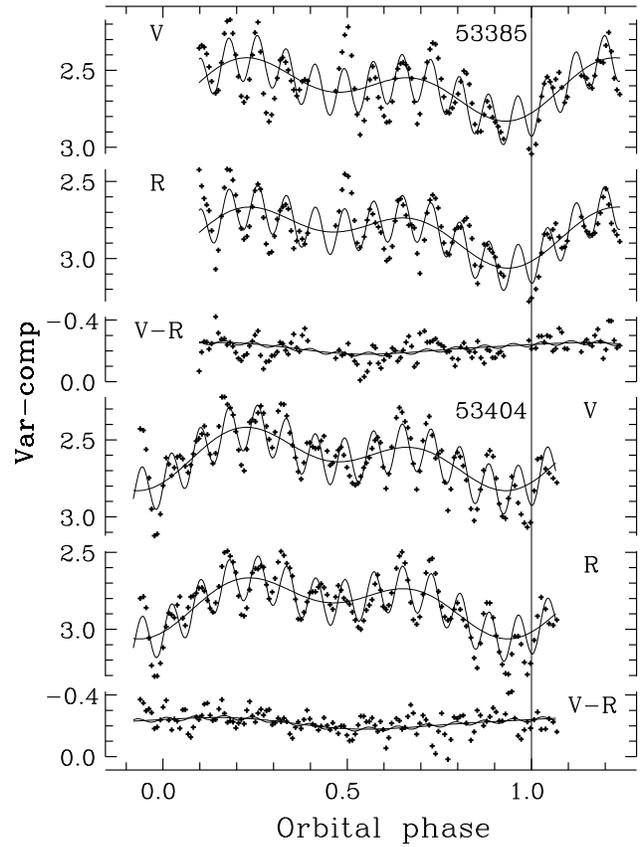}
\label{f4}
\caption{Brightness and color variations for
original and interpolated points (dots) and approximations of the
orbital and ``orbital+spin" variability.}
\end{figure}

De Martino et al. (1995) have applied a multi-sinusoidal fit
assuming multi-harmonic shape of the orbital light curve with
frequencies $\omega_J=j\Omega,$ where $\Omega=1/P$ is the
frequency corresponding to the orbital period $P,$ $j=1..s.$ In
addition, they took into account the 913-sec pulses with frequency
$\omega_{s+1}$ assuming its nearly sinusoidal shape.

The mathematical model for this two-frequency process is
\begin{equation}
I(t)=I_0+\sum_{j=1}^s A_j\cos(\omega_j(t-T_{0j})).
\end{equation}
Here $I(t)$ is intensity (computed from the magnitudes) and $A_j$
are semi-amplitudes of variations, and $T_{0j}$ are initial epochs
of the maxima of the waves corresponding to frequencies
$\omega_j.$ We use fluxes instead of magnitudes, to make results
comparable to those of de Martino et al. (1995). According to the
calibration (cf. Johnson 1955, Allen 1973), the magnitude
V=15\fm00 corresponds to the value of the monochromatic flux
$f_{\lambda}=10^{-14.44}=3.63\cdot 10^{-15}$ erg
s$^{-1}$cm$^{-2}$\AA$^{-1}$).

Results of the fit for the value $s=6+1$ (for orbital and spin
variability, respectively), which had been adopted by de Martino
et al. (1995), are listed in  Tables 2 and 3. It should be noted that
the parameter $I_0$ generally differs from the sample mean value
$\langle I\rangle.$

For two short runs 52635 and 52638, the distribution of
observations in the orbital phase is bad, thus the values of the
parameters are highly biased and thus have no physical meaning,
despite the fit itself is being good enough. The values deviating from
zero less than $3\sigma$ are not statistically significant, and
marked by the symbol ":" as "bad" ones. For the first run, all
values are thus marked as "bad", despite the amplitude for the
spin variability having a reasonable value. For other nights, the
estimates of the mean semi-amplitude of the spin variations varied
in the range $(0.38-0.46)\cdot10^{-15}$ erg
s$^{-1}$cm$^{-2}$\AA$^{-1}.$ This interval slightly exceeds the
error estimate, indicating cycle-to-cycle and thus night-to-night
variability of this parameter.

One may note that the ratio $P_{orb}/P_{spin}=12.74$ is far
from an integer value, thus the orbital light curve differs from
that of the next cycle by a spin phase shift of a quarter of
period.



For all nights, the amplitudes of the waves with the orbital
period $R_1,$ its first harmonic $R_2$ and the spin period $R_7$
are statistically significant. The value $R_3$ slightly exceeds
$3\sigma$ only for two nights. For other nights, as well as for
higher harmonics of the orbital period, the amplitudes $R_3,$
$R_4,$ $R_5,$ $R_6$ are not statistically significant, and
generally should not be taken into account. However, we present
results in Tables 2 and 3 for the same $"6+1"$ multi-sinusoidal
model, as has been proposed by de Martino et al. (1995).
The multi-sinusoidal fits for fluxes are shown in Fig.3.
The fits show the presence of some variations of the orbital light
curve, if taking into account such a large number of harmonics.

\begin{figure}
\psfig{file=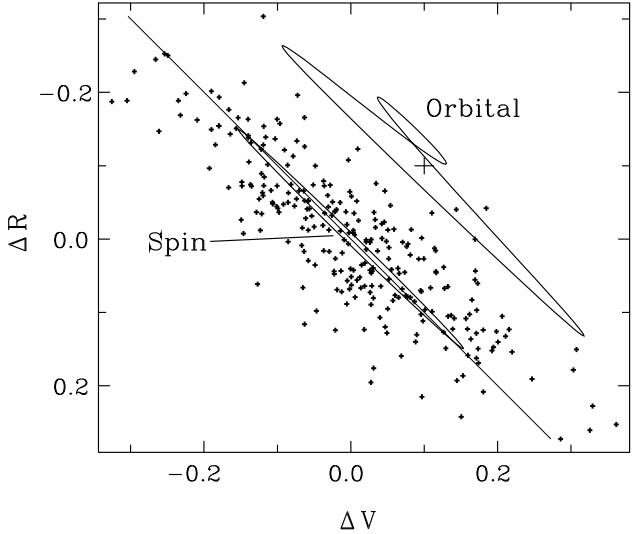}
\label{f5}
\caption{Two-color diagram for the
deviations of original and interpolated points from orbital fit
(dots). The line corresponds to the unitary slope $\Delta R=\Delta
V.$ The closed lines correspond to the orbital fit (shifted by
$+0\fm1$ in V and $-0\fm1$ in R) and deviations of the
``spin+orbital" fit from the pure ``orbital" one. The orbital curve is shifted by $+0\fm1,-0\fm1$ for illustrating purposes. The cross indicates $\pm1\sigma$ values for accuracy of the smoothed curves in V and R.}
\end{figure}

We may conclude, that, for our VR observations obtained in 2002-2005,
the highest harmonics are not significant, contrary to the B observations
of de Martino et al. (1995) obtained in 1983-1992. For firm
conclusions, new multi-color observations of the object are
needed.

The direct comparison of the fluxes between this work and that by
de Martino et al. (1995) is not possible, as they have published
their results in another photometric system (B). The peak-to-peak
amplitude (Table 3) of the multi-sinusoidal fit of the orbital variability
for 5 runs, which completely covers the period, varies from 0\fm38
to 0\fm46 in V. However, for the same nights, the amplitudes in V
and R are practically the same around 0\fm45.

Despite a generally good agreement of the multi-sinusoidal fit
with the observations, showing both a-sinusoidal variability of
the orbital light curve and the spin pulses superimposed, there
are some important differences to the "running sine" fits. In
the "global"  multi-sinusoidal model, the amplitude of the spin
pulses is constant (in the "flux" representation), thus its
apparent "magnitude" variability with phase is caused by the
variability of the smoothed phase dependence of the orbital flux,
and not by the physical variability of individual spin pulses.

\subsection{Night-to-night variability}
The mean brightness variations during 9 nights of our observations show only minor variations with a peak-to-peak amplitude of 0\fm10, as one may see from the values of $I_0$ in Table 2. This value is smaller than that 0\fm22 reported by McHardy et al. (1984). This may be an effect of a small sample. However, if assuming that these variations are more pronounced in the ultraviolet, and given that McHardy et al. (1984) have used unfiltered observations, one may interpret such a difference in amplitudes as a result of different photometric systems. Unfortunately, we have not found other amplitude estimates in the literature, thus the question is still open.

\section{Color variability with the phase of orbital and spin period}

\subsection{Quasi-Simultaneous VR Observations}

Our observations in V and R have been made using alternatively
changing filters. In this case, direct color measurements are
not possible, as BG CMi shows fast variability. To make a set
of quasi-simultaneous observations, the interpolating or smoothing
technique is to be applied. For another cataclysmic variable TT
Ari, we have used the
method of ``running parabolae" Andronov (1997) to make a set of
simultaneous smoothed values in 3 colors UBV (Tremko et al. 1996,
Andronov et al. 1999). That method is
effective for stars with flickering-dominated variability.

For BG CMi, the major amplitude of variations is due to orbital
and spin variability, thus we have applied an interpolating
procedure to determine brightness at missing times using
times of observations in another filter. For the local sequence VRVRVRV, we have applied a local
cubic polynomial fit using two nearby observations before and
after. For regularly spaced observations, this will lead to an
interpolating formula
$\tilde{x}_i=(9(x_{i-1}+x_{i+1})-(x_{i-3}+x_{i+3}))/16$ assuming
that indices $i-2, i+2$ correspond to the same channel as the
point $i.$ The statistical error of this smoothing value is
$\sigma_{\tilde{x}}=(41/64)^{1/2}\sigma_x\approx0.80\sigma_x,$
where $\sigma_x$ is an accuracy estimate of the individual signal
value. Such an approximation is better than a linear one in our case
of relatively fast oscillations. For real observations with nearly
the same time interval between subsequent points, the weight
coefficients have been computed according to a polynomial
interpolation. Near the borders, where the point $i\pm 2$ is
located outside a run of equidistantly spaced times, we have
applied a parabolic fit using the points $i\mp1,i\pm1,i\pm3$ or a
linear interpolation using the observations $i\mp1,i\pm1.$ No
extrapolation has been made to avoid large errors.

As a result of this procedure (see Andronov and Baklanov (2004)
for a detailed study of statistical properties of such
approximations), we have obtained 269 pairs of ``simultaneous" VR
observations, for which the instrumental color index V-R has been
computed. The data points which are present only for one color for a trial time, have been excluded.

As unfortunately the brightness of the comparison stars
in R is unknown, we cannot convert instrumental magnitudes and the
corresponding color index to the standard system. Thus the
magnitudes will be expressed as Var-Comp, and the colors will be
represented with respect to the comparison star, i.e. our
V-R=(V$_{var}$--V$_{comp}$)--(R$_{var}$--R$_{comp}$).

\subsection{Brightness and color variations}

The corresponding light and color curves for this set of
``original+interpolated" points are shown in Fig. 4.
Despite cycle-to-cycle variability of the individual spin pulses, their characteristic times are much longer than the time resolution of our observations, thus the interpolation has not  biased the shape of the light curve, as one may see by comparing with the data presented at high-harmonic fits for intensities at Fig.3. Because of the relatively small amplitude of orbital variations $(\sim 0\fm4),$ the difference in the fits of magnitudes and intensities is compared with the error estimates, and may only be important at the grazing eclipse and subsequent dip.

The degree of the trigonometric polynomial for the multi-sinusoidal fits have been reduced to $s=2+1,$ taking into account that higher harmonics of the orbital variability are not statistically significant (see Table 2).
Despite the long runs are presented in Fig. 4 separately, the fits have been obtained for a long joint run to decrease the night-to-night scatter. Thus the approximation of orbital variability is the same for both nights. The spin frequency is 94.58 cycles/day or 12.7447 cycles per orbital period, thus the phase of the spin variations changes by a quarter of a period every $P_{orb}.$ However, in our observations, the interval between the orbital minima was apparently $141P_{orb}=1797.003P_{spin},$ so we have a rare event of coincidence of both spin and orbital phases in both nights, so the fits for the runs are the same.

Under such conditions, all differences between the curves are due to physical variability of the star and not to the effects of dependence on two independent phases. The high-amplitude spin pulsation cycle at $\varphi_{orb}\sim0.5$ seen in the run 53385 has no counterpart for the run 53404, where in the vicinity of this phase an opposite phenomenon (amplitude decrease) have been observed.

The mean magnitudes with respect to the comparison star are $m_{0V}=2\fm612(7)$ and $m_{0R}=2\fm830(6)$, so the relative color index is equal to $m_{0V}-m_{0R}=-0\fm218(4)$ and corresponds to a slightly larger temperature than the comparison star. This color difference is small enough to justify use of the star as the comparison one.

The orbital light curve has two minima separated by a half-period. If the phase zero corresponds to the grazing eclipse, the secondary minimum may be interpreted by a partial eclipse of the illuminated secondary by the disk and/or ellipticity of the disk. The semi-amplitudes of the contribution with the orbital period are $R_{1V}=0\fm116(10)$ $R_{1R}=0\fm123(9),$ with nearly the same values for its harmonic: $R_{2V}=0\fm119,$ $R_{2R}=0\fm114(8).$

To determinine effective amplitudes in other types of cataclysmic variables with aperiodic light curves, Andronov et al. (1999) used the r.m.s. deviation $\sigma_*$ of the smoothed values from the mean. For periodic variables, one may use an effective semi-amplitude $R_*=(R_1^2+R_2^2)^{1/2}=2^{1/2}\sigma_*.$ For our data, such values are practically identical for two colors: $R_{*V}=0\fm166(10)$ and $R_{*R}=0\fm167(9).$ Such behaviour of the intermediate polar BG CMi is intermediate between an increase of amplitude with {\em decreasing} wavelength in non-magnetic nova-like variables with superhumps (e.g. MV Lyr, Walker (1954), TT Ari,  Andronov et al. (1999)) and with eclipses (e.g. DW UMa, Ostrova et al. (2005)) or with {\em increasing} wavelength in magnetic classical polars (e.g. AM Her, Szkody \& Brownlee (1977)).

The multi-sinusoidal analysis of the color index shows that the amplitude corresponding to the orbital period $(R_1=0\fm036(6))$ is much larger than for other frequencies. The star becomes most blue at the orbital phase 0.42 and most red at 0.82.

A similar situation is with the semi-amplitudes of the spin variations: $R_{3V}=0\fm154(9)$ and  $R_{3R}=0\fm147(8),$ so the color excess of the spin variations $\Delta V-R=-2.5\lg(R_{3V}/R_{3R})=-0\fm5(9)$ is zero within error estimates. Nearly the same similarity of the spin amplitudes is observed in another intermediate polar FO Aqr  but with much smaller values $R_{3V}=0\fm087(6)$ and  $R_{3R}=0\fm095(4)$ (Andronov et al. 2005b).

\subsection{Two-color diagram}

The two-color diagram for the deviations of the data from the orbital fit is shown in Fig. 5. The line of equal shifts $\Delta R=\alpha\Delta V$ satisfactorily represents individual points with $\alpha=1,$ and is in excellent agreement with the sinusoidal component. As there is a very small difference between the amplitudes discussed in previous section, the slope $\alpha$ seems to be slightly less than unity. However, this effect is within the error estimates.
Also the phase shift between oscillations in V and R is not statistically significant, thus making the minor axis of the ellipse vanish. There is a scatter of individual points of 0\fm048, which is much larger than the accuracy estimate of the individual measurement of the star of such brightness (0\fm01, see Fig.1). Such scatter may correspond to a possible flickering with much higher amplitude in V than in R, that is usual for accretion. This hypothesis may be supported by a the result that the $V-R$ color is a much more high correlated with $V$ than with $R$. The corresponding correlation coefficients are $r[V-R,V]=0.5(5)$ and $r[V-R,R]=0.20(8)$, respectively. This may be also the result of the contribution of ``strongly colored" spin cycles, when the color variations distinctly deviate from the smooth orbital curve. However, based on the present material, it is not possible to obtain sufficient statistics on the occurence of ``colored" and ``non-colored" pulses.

\begin{figure*}
\psfig{file=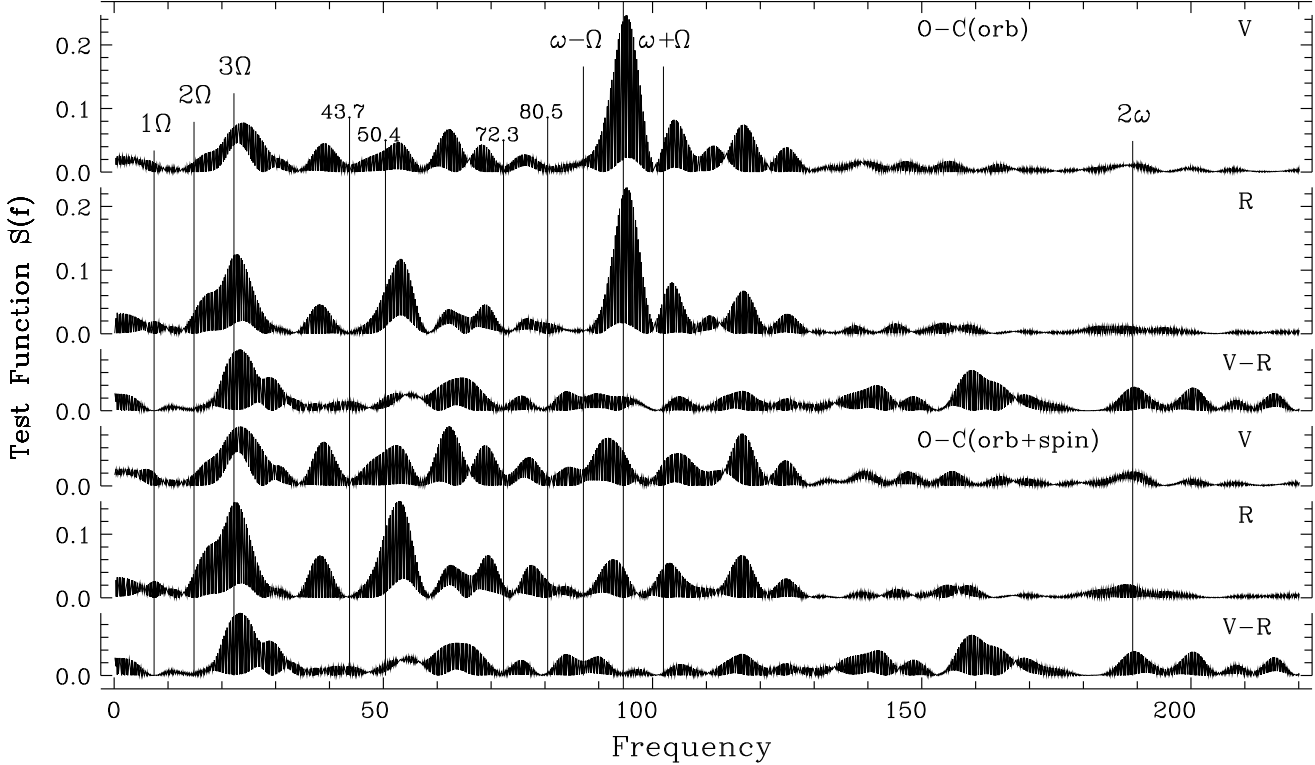}
\label{f6}
\caption{Periodograms $S(f)$ for
deviations of the brightness and color from the orbital fits for
two nights of two-color photometry. Vertical lines mark positions
of the orbital $(\Omega)$ frequency and its harmonics, as well as
of the spin $(\omega)$ frequency and its harmonic $(2\omega)$ and
sidebands $(\omega\pm\Omega)$. Numbers correspond to the
frequencies suspected by Hellier (1999) and Paterson and Thomas (1993).}
\end{figure*}

Optical, UV and X-Ray fluxes of intermediate polars show highly structered accretion geometry (cf. Kim \& Beuermann 1995, 1996). Our results can be used to model such an accretion flow.

The difference in the shape of the orbital light curves in two colors leads to a change of the trajectory at the two-color diagram from a line to an ``8-like" figure.
Such difference in behaviour is the observational basis for future theoretic modeling of the orbital variability.

\subsection{Periodogram analysis}
To study differences in light curves in different colors, we have applied the periodogram analysis to the deviations of the individual data points from the ``orbital" $O-C_{orb}$ and ``orbital+spin"  $O-C_{orb+spin}$ fits. For some objects, there are peaks present at beat frequencies, which are the combinations of the orbital frequency $\Omega$ and the spin frequency $\Omega$ (cf. Warner 1986, 1995, Wynn and King 1992, Patterson 1994, Hellier 2001).

The test-function $S(f)$ has been used (see Andronov (1994) for details). The deviations  $O-C_{orb}$ show the series of strong peaks corresponding to the spin frequency and its 19-day aliases (the interval between the runs 53404 and 53385). The peaks near this frequency practically disappear for $O-C_{orb+spin}$, and there is no evidence for the spin frequency for the color index for both types of deviations. However, there are strong peaks near $3\Omega$ (but not exactly ``at"). Also in R there is an apparent peak at $\sim55$ cycles/day. We suggest that these apparent peaks arise from instabilities of the light curves.
Such quasi-periodic oscillations with a time-scale of 15-60 minutes are characteristic for some cataclysmic variables of similar orbital periods (e.g. Tremko et al. 1996, Andronov et al. 1999). However, our periodograms show no significant peaks at frequences, where the peaks have been found in previous years by Patterson and Thomas (1993) and Hellier (1999). There is no evidence for significant peaks at the beat/multiple frequencies, where such peaks are seen in other cataclysmic variables.

\subsection{Scalegram analysis}

The scalegram analysis has been carried out using the ``running
parabola" approximation. The $``\sigma(\Delta t)"$ scalegram
(Andronov 1997) shows a shape characteristic for a two-component
signal. For such a case, is useful the $``\Lambda(\Delta t)"$
scalegram (Andronov 2003), which is shown in Fig. 7 for 2002-2004, when one-color photometry provided better time resolution. The most
prominent maximum corresponds to the spin variability with an
effective period of $P_{spin,eff}=0\fd01058$ and semi--amplitude
$r_{spin,eff}=0\fm139.$ This value slightly exceeds the mean semi-amplitude of the spin variations obtained from multi-sinusoidal fits $r_{ms,eff}=-2.5\lg(((I_0+I_7)/(I_0-I_7))^{1/2}),$ which is equal to 0\fm127. This excess argues for more rapid variations of the individual cycles and their phase shifts. For a pure sinusoidal signal, both methods should give closer results.

The second peak at the $``\Lambda(\Delta t)"$ scalegram corresponds to the first
harmonic of the orbital variability with $P_{2,eff}=0\fd061$ and
$r_{2,eff}=0\fm083.$ The third peak in between may correspond to
the third harmonic of the orbital period, because the shape is not
sinusoidal. As expected, the effective values for the spin period
coincides much better with the values obtained by other methods
than that for the orbital variability. This is caused either by
the duration of observations comparable with the orbital period
or by the non-sinusoidal shape of the orbital variability.

\section{194-minute orbital period}
\subsection{Orbital minima and dips}
The orbital minimum has a specific asymmetric shape at its bottom,
showing a narrow dip occurring by $\sim 0.07P$ later than the
mid-eclipse, as one may see from Fig.6 of Patterson and Thomas
(1993). To study further period behavior, we have used 42 minima
published in Table 1 of their paper. It seems that the one
corresponding to HJD 2445730.443 seems to be a misprint. Following
the authors, we have checked the original reference of Singh et
al. (1991) for the orbital {\sl maximum} and corrected the decimal
part of this point to 0.120. Then we have collected 5 minima
obtained by Pych et al. (1996). In this work, we have determined 3 moments of centers of minima
from the function $a(t_0)$ obtained using the ``running sine" fit
for the nights 53035, 53036, 53052, when the coverage of eclipses
was good. The moments of dips for these nights have been determined
as those corresponding to the minumum of $a(t_0).$ For the nights
53384, 53404, the coverage of eclipses is not complete, so we have
used the multi-periodic fit $(s=6+1).$

Additionally, we have analyzed 3 nights of unpublished unfiltered CCD photometry with a 120-sec integration time, which was obtained by Foote (2003) at the 93cm telescope in Utah as a part of activity of the Center for Backyard Astrophysics \mbox{(http://cba.phys.columbia.edu).}
From 3 nights, 4 mean minima have been determined using multi-sinusoidal ``(2+1)" fits, one of them (52673.6927) corresponds to the mean light curve for all these nights. This "white light" photometry is marked as ``W" in the text.

All orbital minima determined in this work are listed in the Table 4.

Our minima show smaller scatter than that from the tables by
Patterson and Thomas (1993) and Pych et al. (1996),
where the phases of minima range from $-0.13$ to $+0.22,$
because of the inhomogeneity of their collection of previously
published (or even estimated from the light curves) minima.
Adding our 11 points to their 47 minima, the best fit ephemeris
has been obtained
\begin{eqnarray}
Min.HJD&=&2445020.3894+0.13474841\cdot E\nonumber\\
&&\hspace{0mm}\pm\hspace{10.5mm}20\hspace{3mm}\pm\hspace{10mm}6\\
Min.HJD&=&2448368.7225+0.13474841\cdot (E-24849)\nonumber\\
&&\hspace{0mm}\pm\hspace{10.5mm}12\hspace{3mm}\pm\hspace{10mm}6
\end{eqnarray}
The second ephemeris corresponds to the minimum error estimate of the initial epoch, which is closest to a sample mean of times used.

From our minima timings, we may estimate phase difference between
the mid-eclipse and orbital dip to range from 0.04 to 0.11P.
The mean phase of our 7 VR minima with respect to the ephemeris of
Patterson and Thomas (1993) is
$\langle\varphi_{min}\rangle=-0\fp056\pm0\fp007;$ it reaches
$8\sigma$ and thus is to be taken into account. Similar negative O-C are seen for the ``W" observations obtained by Foote (2003). According
to the new ephemeris (6), the mean phase
$\langle\varphi_{min}\rangle=-0\fp012\pm0\fp007$
is much close to zero, but there is still some deviation.
It may be explained by inhomegeneity of the previously
published data, where both orbital minima and dips have been
included without separation. The dips occur at the mean phase
$\langle\varphi_{dip}\rangle=+0\fp049\pm0\fp013$,
according to the ephemeris (6), so the mean phase difference is
$\langle\varphi_{dip}\rangle-\langle\varphi_{min}\rangle=+0\fp062\pm0\fp015.$
There is also a suggestion that the minima in R occur by
$\sim0\fp034\pm0\fp007$ later than in V. However, it is based
only on two nights of two-color observations, when the minima
were badly covered by the observations, thus it may have resulted
from statistical fluctuations of observational errors.
This suggestion has to be checked in future studies.

The peak-to-peak
amplitudes of the orbital variations, obtained from the multi-periodic fit, are $0\fm43$ and  $0\fm41$ for the runs V\,53385 and V\,53404, respectively. For the filter R, these values are practically the same: $0\fm43$ and  $0\fm41.$

As the orbital periods of cataclysmic variables undergo evolutionary decrease (cf. Patterson 1984), we have checked the O-C diagram for the presence of a quadratic term in an ephemeris. Unfortunately, the scatter of the early data compiled by Patterson and Thomas (1993) is rather large (r.m.s. value of 0\fp069). So, despite much better accuracy of the data in their and our work and that by Pych et al. (1996), the obtained value of the $\dot{P}_{orb}$ deviates from zero only by $0.9\sigma,$ and thus the hypothesis of decrease of the orbital period may not yet be justified from all available data spanning more than two decades.

\subsection{Phase dependence of the pulse-averaged brightness and amplitude}

\begin{figure}
\psfig{file=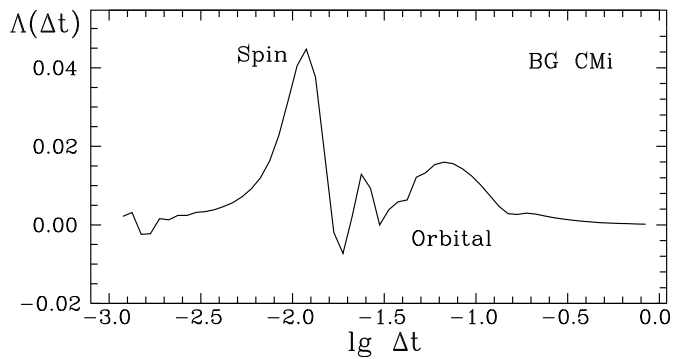}
\label{f7}
\caption{The dependence of the test function $\Lambda(\Delta t)$ on the filter half-width  $\Delta t.$ The two highest peaks correspond to the spin and the harmonic of the orbital variability.}
\end{figure}

The characteristics of the ``running sine" fits are shown in Fig.
8. The most important of them are the local mean (pulse-averaged
brightness) $a(t)$ and semi-amplitude $r(t)$ of the spin pulse.
The orbital variability is prominent with the broad $(0.2P)$
minimum and the lagged dip. The amplitude shows night-to-night
scatter being relatively constant at the phase intervals
0.45--0.50. The phases have been computed according to the
ephemeris (4) by Patterson and Thomas (1993).

At the phases of orbital minimum, the semi-amplitude shows a
minimum for all nights. However, the dip is characterized by an
abrupt peak of the semi-amplitude. The peak lags the dip by
$\approx0.02P,$ that may be partially explained by an abrupt
ascending branch after the dip causing an {\em apparent} increase
of the amplitude. However, this phenomenon may not explain the
peak completely, as the amplitude really increases, as one may
check the original and smoothed light curves.

\section{Variability of the spin period}

\subsection{Determination of times of extrema}

\begin{figure}
\psfig{file=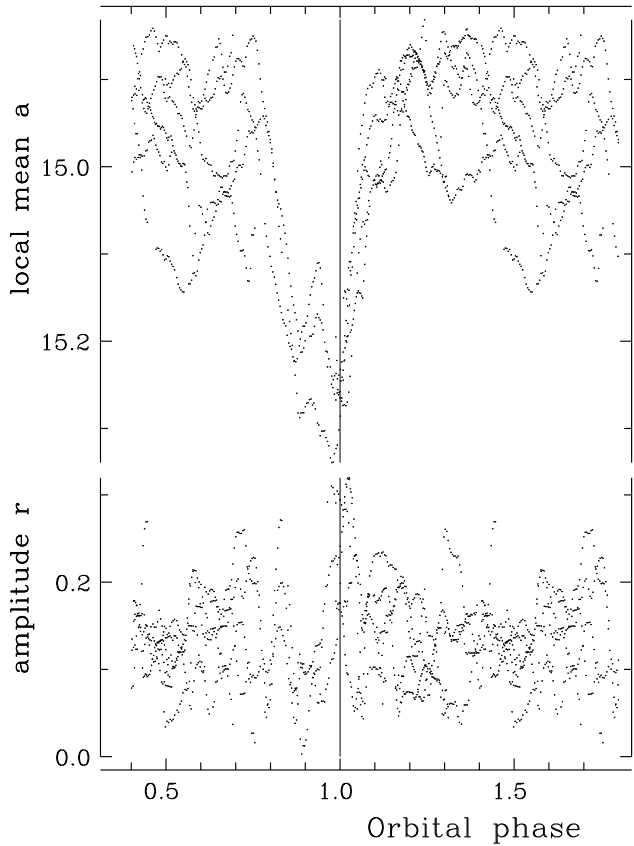}
\label{f8}
\caption{The dependence on the orbital phase of the local mean and semi-amplitude of the spin variability.}
\end{figure}

\begin{figure}
\psfig{file=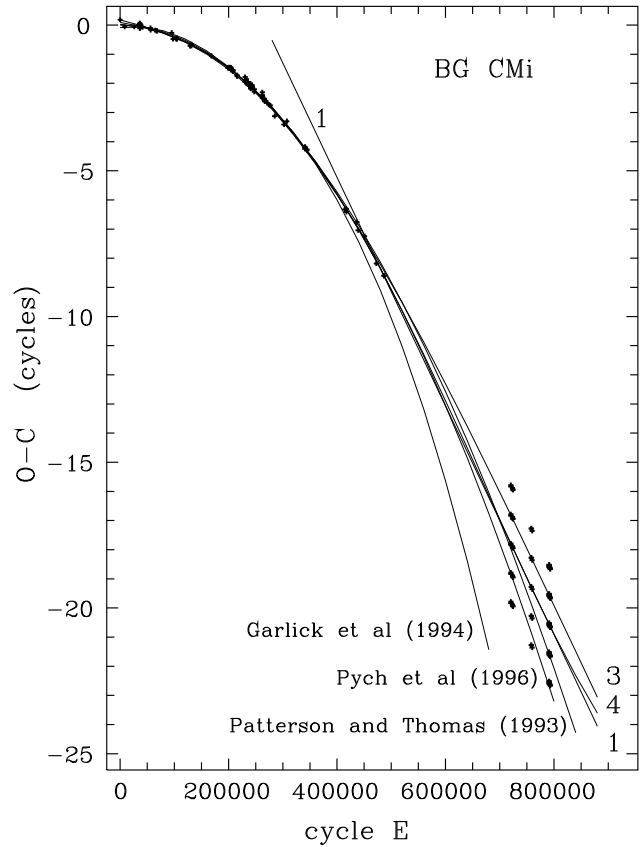}
\label{f9}
\caption{O-C values for times of spin maxima calculated with the linear ephemeris (Pych et al. 1996) with a cycle numbering corresponding to the best 4-th order polynomial fit. Fits are shown, which correspond to ephemerides published by Patterson and Thomas (1993), Garlick et al. (1994) and Pych et al. (1996) for polynomials of order 2, 3 and 2, respectively. The order of our polynomial fits with optimal cycle counting are marked by numbers. Our parabolic fit is close to that of Pych et al. (1996) and is not shown separately. Because of possible cycle miscount after the gap, our data are shown with shifts of integer number of periods.}
\end{figure}
For the analysis of one-color observations in 2002-2004, we have used the method of ``asymptotic parabolae" (AP), proposed by Marsakova and Andronov (1996) for timings of extrema of signals with cycle-to-cycle variability. The tables of 58 maxima and 63 minima have been published separately by Andronov et al. (2005a). They have used these results to obtain 5 nightly and 2 seasonal ``mean" maxima timings with much better accuracy. The extrema timings show a scatter of phases within a relatively large ``corridor" of $\pm0\fp3,$ except 3 points deviating from the mean value toward even larger values. These 3 points have been omitted. The mean deviation of the minima from the phase 0.5 corresponding to maxima is only $+0\fp003\pm0\fp020$ and thus is not statistically significant. This is in a contradiction to a qualitative suggestion of more abrupt brightness increase than decrease (de Martino et al. 1995).

For the observations obtained in 2005, the use of alternating filters had led to half the number of data per cycle, and thus it was not possible to apply the same method, which is effective to study characteristics of individual cycles. For determination of the individual extrema in this case, we have used the multi-sinusoidal fit (5) for ``Var-Comp" magnitude differences.
Because of the statistical symmetry of spin pulses, we have used only the wave with a spin period without any additional harmonics, contrary to fitting the shape of the orbital phase curve.
These timings for earlier one-color observations are in good agreement with the results presented by Andronov et al. (2005a), but, for uniformity, we have used results obtained using the multi-sinusoidal fits.

The corresponding timings of the ``mean" maxima are listed in Table 3. To obtain more accurate results, we have  determined ``mean" maximum for all 3 nights of the ``W" observations, as well as for two long runs in V and R.


\begin{table}
\label{t3}
\caption{Characteristics of the multi-sinusoidal fit of individual
runs of observations of BG CMi (continuation of Table 2).}
\begin{center}
\begin{tabular}{ccccc}
\hline
$T_{07}-2400000$&$\langle I\rangle$&$R_7$&$\Delta m$\\
\hline
V 52635.32067$\pm$0.00047& 3.492 &0.320:$\pm$0.126\\
V 52638.30276$\pm$0.00022& 3.861 &0.444~$\pm$0.056\\
V 53035.10202$\pm$0.00020& 3.536 &0.457~$\pm$0.055&0\fm46\\
V 53036.19093$\pm$0.00021& 3.832 &0.385~$\pm$0.048&0\fm39\\
V 53053.02230$\pm$0.00018& 3.597 &0.433~$\pm$0.047&0\fm38\\
V 53383.09882$\pm$0.00777& 3.508 &$0.284:\pm$0.176\\
V 53384.14518$\pm$0.00057& 3.976 &$0.802:\pm$0.297\\
V 53386.06917$\pm$0.00037& 3.524 &$0.561\pm$0.126&0\fm45\\
V 53405.04680$\pm$0.00043& 3.440 &$0.467\pm$0.117&0\fm44\\
R 53383.09798$\pm$0.00078& 277750&$34337\pm$14845\\
R 53384.14560$\pm$0.00067& 311765&$57053\pm$22282\\
R 53386.06915$\pm$0.00036& 276381&$41497\pm$8917&0\fm47\\
R 53405.04708$\pm$0.00044& 267748&$34647\pm$8946&0\fm43\\
\hline
\end{tabular}
\end{center}
\end{table}

\begin{table}
\label{t4}
\caption{Moments of orbital minima and dips}
\begin{tabular}{rrrr}
\hline
Min BJD&E&O-C&type\\
\hline
W 52669.9227 & 56769& -0.0046& Mid-eclipse\\
W 52670.7231 & 56775& -0.0127& --"--\\
W 52679.7595 & 56797& -0.0075& --"--\\
W 52673.6927 & 56842& -0.0044& --"--\\
V 53035.0875 & 59479& -0.0085& --"--\\
V 53036.1663 & 59487& -0.0077& --"--\\
V 53053.0073 & 59612& -0.0102& --"--\\
V 53386.1087 & 62084& -0.0074& --"--\\
R 53386.1122 & 62084& -0.0039& --"--\\
R 53404.9759 & 62224& -0.0050& --"--\\
V 53405.1050 & 62225& -0.0106& --"--\\
V 53035.1023 & 59479&  0.0063& Orbital dip\\
V 53036.1724 & 59487& -0.0016& --"--\\
V 53053.0170 & 59612& -0.0005& --"--\\
V 53405.1152 & 62225& -0.0004& --"--\\
\hline
\end{tabular}
\end{table}

\begin{table}
\label{t5}
\caption{Times of mean maxima from the multi-sinusoidal fit of individual
runs of observations of BG CMi.}
\begin{center}
\begin{tabular}{ccccc}
\hline
E&$t_{max}-2400000$&$\phi_5$&$\phi_{Pych}$&Rem\\
\hline
720253& 52635.32067 (47)&-0.05&-17.83&V\\
720535& 52638.30276 (22)& 0.01&-17.78&V\\
723524& 52669.90440 (37)& 0.03&-17.88&W\\
723608& 52670.79209 (19)&-0.00&-17.92&W\\
723898& 52673.85825 (15)& 0.01&-17.92&W\\
724460& 52679.80003 (21)& 0.01&-17.95&W\\
758066& 53035.10202 (20)& 0.02&-19.27&V\\
758169& 53036.19093 (21)& 0.02&-19.28&V\\
759761& 53053.02230 (18)& 0.00&-19.36&V\\
790981& 53383.09798 (78)&-0.04&-20.60&R\\
790981& 53383.09882(777)& 0.04&-20.52&V\\
791080& 53384.14518 (57)& 0.01&-20.56&V\\
791080& 53384.14560 (67)& 0.05&-20.52&R\\
791262& 53386.06915 (36)&-0.01&-20.59&R\\
791262& 53386.06917 (37)&-0.01&-20.58&V\\
792203& 53396.01793 (10) &-0.02&-20.62&V\\
792203& 53396.01806 (10) &-0.01&-20.61&R\\
793057& 53405.04680 (43)&-0.03&-20.67&V\\
793057& 53405.04708 (44)&-0.00&-20.64&R\\
\hline
\end{tabular}
\end{center}
\end{table}

\begin{table}
\label{t6}
\caption{Seasonal ephemerids for spin maxima of BG CMi according to the linear fit $t_E=T_0+P\cdot(E-E_0).$}
\begin{center}
\begin{tabular}{cccc}
\hline
$T_0-2400000$&$P$&$\phi_5$&$E_0$\\
\hline
52661.32970(11)&0.010572652~(66)&~0.001  &722713\\
53041.43489~(2)&0.010572451~(28)&~0.013 &758665\\
53390.87981~(9)&0.010572466(116)&-0.003 &791717\\
53105.31448~(6)&0.0105725772(20)&~0.020&764707\\
\hline
\end{tabular}
\end{center}
\end{table}


\subsection{Models of variability}

The variability of the spin period of BG CMi had been detected by
Singh et al. (1991) and Augusteijn et al. (1991). Patterson \& Thomas (1993) determined a quadratic ephemeris based on their own and compiled times of maxima. Garlick et al.
(1994) suggested even a third-order polynomial to fit the O-C
diagram. In more recent study, Pych et al. (1996) rejected this
suggestion, arguing for a sufficience of the second-order
polynomial.

For reference, we have used again the linear part of the most recent ephemeris (3)
published by Pych et al. (1996). The long-term variability of the period and thus the initial epoch is negligible for such a {\sl local} fit, contrary to the complete interval.

In Fig. 9, different models of O-C variations are shown. We checked the ephemeris by Pych et al. (1996).
Despite it satisfactory fits to the data obtained in 2002-03 and 2005, the spin maxima in 2003 occurred out of phase, so such a fit is not acceptable. The ephemeris by Patterson \& Thomas (1993) leads to a shift of the cycle numbering of all our data by unity, thus has the same problem with the year 2004. All our data may be well fitted by the ephemeris by Garlick et al. (1994) assuming a numbering shift of 7 cycles. However, it badly fits the data by Pych et al. (1996).
Thus no fit from previously published ones describes the whole data set.

After such a long gap of 6 years, the main problem is the correct cycle numbering. For this purpose, we have determined the best values for separate seasons, using the maxima timings from Table 5, and the preliminary cycle numbering from the ephemeris by Pych et al. (1996). They are listed in Table 6.
The weighted mean value had allowed us to obtain a corrected difference of the cycle numbers between the initial epochs for the first (2002-03) and last (2005) seasons.

Assuming the new linear ephemeris for all our data (Table 6), one has to change the initial ephemeris by Pych et al. (1996). Assuming the quadratic model for O-C, from the values at the period $E=0$ and at $E=764707$ (our epoch for 2002-2004), we get $Q=-2.71(2)\cdot10^{-13},$ significantly different from the value published by Pych et al. (1996). This shows that the ``global" parabolic fit for all data is not satisfactory. Assuming an additional cubic term $Q_3E^3$ with fixed values of other 3 parameters, the value of $Q_3=+9.7\cdot10^{-20}$ had been estimated assuming that, for a negative $\dot P,$ the second derivative $\ddot P$ is positive. This indicates a deceleration of the spin-up of the white dwarf. This result is based only on the analysis of period values at the beginning and end of known data, and is not dependent on possible minor cycle miscounts.

Our linear ephemeris fits well the last timings of Pych et al. (1996), so formally one may apply a composite ``spline-type" model for the O-C with parts of spin-up and constant periods. It resembles the ``asymptotic parabola" model (Marsakova and Andronov 1996) for extrema timings.
Such abrupt period changes are often observed in classical eclipsing binary stars (e.g. Kreiner et al. 2001). Another approach is to apply high-order polynomials (e.g. Kalimeris et al. 1994).

For such polynomial models, we have chosen a free parameter which corresponds to the cycle number $E_f$ of one of the minima from our sample. We have chosen the initial epoch for the last season 2005. The initial cycle number was chosen to correspond to the quadratic ephemeris by Pych et al. (1996), i.e. 791717. Then it was modified within limits of $\pm10$ cycles. For each trial value of $E_f$  and of the degree of model, the following iterations were made.
At first, this point replaced all our 19 timings, so giving high weight for this point.
The computed best fit polynomial ephemeris has been used to determine the cycle numbers of all 19 timings. The numbering of the previous 90 points remained the same. Then a final polynomial fit was obtained. The test function was the r.m.s. deviation of the phases from the fit $\sigma_{O-C}.$

For the parabolic fit, the numbering remained the same, showing unique deep minimum at the $\sigma_{O-C}(E_f)$ dependence at $E_f=791717.$ However, the maximal deviations exceded 0.4P, thus the fit is not acceptable.
The cubic fit has an optimal solution for $E_f=791714,$ which corresponds to the ephemeris
\begin{eqnarray}
Max. HJD&=&2445020.2791(3)+0.0105730179(37)E\nonumber\\
&&-5.28(13)10^{-13}E^2+2.10(10)10^{-13}E^3
\end{eqnarray}
However, the 4-th order polynomial for $E_f=791715$ gives much better fit to our data, with a scatter smaller by 1.5 times than that corresponding to the cubic fit. Thus it is the statistically optimal fit. The corresponding ephemeris is presented in the abstract. However, the ``spline-type" and polynomial models may also not be ruled out.

For a final determination of the cycle numbering and choice of the appropriate model, one needs to analyse the spin maxima obtained between 1996 and 2002. However, the results of these observations are not published yet.

Future new observations will be crucial for the study of the rotational evolution of the white dwarf in this system, as the ephemerides obtained using different models will diverge in the next half-year or more.


\section{Conclusions}
\begin{itemize}
\item The ephemeris for the orbital minima has been corrected.
\item The amplitude of spin variability has an abrupt maximum at the phase of the orbital dip after the mid-eclipse.
\item The orbital light curves in V and R differ, with the color index having its miminimum (highest temperature) at the phase 0\fp42 and maximum at the phase 0\fp82. The statistically optimal degree of the trigonometric polynomial is 2, unlike the previously published value 6.
\item The spin variations in both colors have the same effective amplitudes and colors, contrary to the non-magnetic cataclysic variables with either superhumps or total eclipses and to the classical polars.
\item The ephemeris for the spin maxima in 2002-2005 has been determined.
\item The rate of the spin-up of the white dwarf has been significantly decreased as compared with previous years.
\item The optimal mathematical model for the spin ephemeris corresponds to a fourth-order polynomial. However, because of the 6-year gap after the previously published observations, other cycle numbering over the gap may not be excluded. It may correspond either to the cubic ephemeris or to the ``spline-like" ephemeris with the spin-up replaced by a constant period rotation.
\item The obtained parameters may be used for comparison with theoretic models of accretion in magnetic cataclysmic variables. The further monitoring is needed for studies of rotational evolution of the white dwarf.
\end{itemize}

\begin{acknowledgements}
The authors are grateful to Jerrold L. Foote (CBA Utah) for sending us his unpublished data,
This work was supported by the Korea Astronomy Observatory and Space Science Institute Research Fund 2003 and was partially supported by the Ministry of Education and Science of Ukraine.
\end{acknowledgements}


\begin{thebibliography}{}
\bibitem[1973]{A73} Allen C.W., 1973, Astrophysical Quantities, The Athlone Press
\bibitem[1994]{A94b} Andronov I.L., 1994, Odessa Astron. Publ., 7, 49 \mbox{(http://il-a.pochta.ru//oap7\_049.ps.z)}
\bibitem[1997]{A97} Andronov I.L., 1997, A\&AS 125, 207
\bibitem[2003]{A03} Andronov I.L., 2003,
ASP Conf. Ser., 292, 391
\bibitem[1999]{A99}Andronov I.L., Arai K., Chinarova L.L., et al., 1999,
  AJ., 117, 574
\bibitem[2004]{AB04}Andronov I.L., Baklanov A.V., 2004, Astron. School Rep., 5, 264
\bibitem[2005]{A05a}Andronov I.L., Kim Y.G., Shin J.-H., Jeon Y.B., 2005, ASP Conf. Ser., 335, 355
\bibitem[2005]{A05b}Andronov I.L., Ostrova N.I., Burwitz V., 2005, ASP Conf. Ser., 335, 229
\bibitem[1991]{A91}Augusteijn T., Schwarz H.E., van Paradijs J.,
1991, A\&A, 247, 64
\bibitem[1998]{C98}Chinarova L.L., 1998, in: J.Dusek, M.Zejda (eds.) Proc. 29th Conf. Variable Star Res., Brno, Czech Republic, 38
\bibitem[1999]{C99}Chochol D., Andronov I.L., Arkhipova V.P., et al.,
1999, CoSka, 29, 31
\bibitem[1995]{D95}de Martino D., Mouchet M., Bonnet-Bidaud J. M., et al.,
1995, A\&A, 298, 849
\bibitem[2003]{F03}Foote J.L., 2003, www.scopecraft.com/Text/Observatory/CBA.htm
\bibitem[1994]{G94} Garlick M.A., Rosen S.R., Mittaz, J.P.D., et al.,
1994, MNRAS, 267, 1095
\bibitem[1999]{H99}Hellier C.: 1999, ApJ, 519, 324
\bibitem[2001]{H01}Hellier C.: 2001, Cataclysmic Variable Stars. How and why they vary, Springer Berlin
\bibitem[1995]{H95} Henden A.A., Honeycutt R.K., 1995, PASP 107, 324
\bibitem[1955]{J55} Johnson H.L., 1955, Ann. d'Ap., 18, 292
\bibitem[1994]{K94}Kalimeris A., Rovithis-Livaniou H., Rovithis P., 1994, A\&A, 282, 775
\bibitem[2004]{KAJ04}Kim Y.G., Andronov I.L., Jeon Y.B., 2004, J. Astron. Space Sci., 21, 3, 191 \mbox{(http://ksss.or.kr/dtp/J200409/kimyonggi.ps)}
\bibitem[1995]{K95} Kim Y., Beuermann K., 1995,  A\&A 298, 165.
\bibitem[1996]{K96} Kim Y., Beuermann K., 1996,  A\&A 307, 824.
\bibitem[2001]{K01}Kreiner J.M., Kim C.-H., Nha I.-S., 2001, An Atlas of O-C Diagrams of Eclipsing Binary Stars, Cracow, Poland
\bibitem[1992]{M92}Massey P., Davis L.E., 1992, A User's Guide to Stellar CCD Photometry with IRAF
\bibitem[1994]{AM94b}Marsakova V.I., Andronov I.L., 1996, Odessa Astron. Publ., 9, 127 \mbox{(http://oap09.pochta.ru)}
\bibitem[1984]{M84}McHardy I.M., Pye J.P., Fairall A.P., Warner B., Cropper M., Allen S.,
1984, MNRAS 210, 663
\bibitem[1992]{N92}Norton A.J., McHardy I.M., Lehto H.J., Watson M.G.,
1992, MNRAS, 258, 697
\bibitem[2005]{O05}Ostrova N.I., Shugarov S.Yu., Andronov I.L., 2005, Ap\&SS, 296, 473
\bibitem[1994]{P94}Patterson J.,
1994, PASP, 106, 209
\bibitem[1993]{P93}Patterson J., Thomas G.,
1993, PASP, 105, 59
\bibitem[1986]{P86}Penning W.R., Schmidt G.D., Liebert J., 1986, ApJ, 301, 881
\bibitem[1996]{P96} Pych W., Semeniuk I., Olech A., Ruszkowski M.,
1996, AcA, 46, 279
\bibitem[1991]{S91}Singh J., Agrawal P.C., Apparao K.M.V., Vivekananda Rao P., Sarma M.B.K.,
1991, ApJ, 380, 208
\bibitem[1997]{S97}Shakhovskoj N.M., Kolesnikov S.V., 1997, IAUC, 6760
\bibitem[1977]{S77}Szkody P., Brownlee D.E., 1977, ApJ, 212, L113
\bibitem[1996]{TAC96} Tremko J., Andronov I.L., Chinarova L.L., et al., 1996, A\&A 312, 121
\bibitem[1954]{W54}Walker M.F., 1954, PASP, 66, 71
\bibitem[1995]{W86}Warner B.,1986,  MNRAS, 219, 347
\bibitem[1995]{W95}Warner B., 1995, Cataclysmic Variable Stars, Cambridge Univ. Press
\bibitem[1992]{WK92}Wynn G.A., King A.R, 1992, MNRAS, 255, 83
\end{thebibliography}
\end{document}